\documentclass[twocolumn,prl,aps,showpacs]{revtex4}

\usepackage{bm}
\usepackage{mathrsfs}
\usepackage{amsmath}
\usepackage{amssymb}
\usepackage{graphicx}
\usepackage{amsfonts}
\usepackage{amsthm}
\usepackage{color}
\usepackage{dcolumn}
\usepackage{txfonts}

\begin{document}

\title{Anomalous decoherence effect in a quantum bath}
\author{Nan Zhao}
\author{Zhen-Yu Wang}
\author{Ren-Bao Liu}
\thanks{Corresponding author. rbliu@cuhk.edu.hk}
\affiliation{Department of Physics and Center for Quantum Coherence, The
Chinese University of Hong Hong, Shatin, New Territories, Hong Kong, China}

\begin{abstract}
Decoherence of quantum objects in noisy environments is important in
quantum sciences and technologies. It is generally believed that different processes
coupled to the same noise source should have similar decoherence behaviors and
stronger noises would cause faster decoherence.
Here we show that in a quantum bath, the case can be the opposite. In particular, we predict that
the multi-transition of a nitrogen-vacancy center spin-1
in diamond can have longer coherence time than the single-transitions,
even though the former suffers twice stronger noises from the nuclear spin bath than the latter.
This anomalous decoherence effect is due to manipulation of the bath evolution via flips of
the center spin.
\end{abstract}

\pacs{03.65.Yz, 76.60.Lz, 03.65.Ud}

\maketitle

Decoherence of quantum objects in noisy environments is of paramount importance in
 quantum sciences and technologies~\cite{ClarkeSCqubit,LaddQC,MazeSensing,BalasubramanianMagnetometry,Zhao_magnetometry}.
In particular, decoherence of electron spins coupled to nuclear spin baths in
quantum dots~\cite{PettaT2,KoppensT2,Greilich06Science,XuPump,Maletinsky09NP,BrunnerHoleSpin,PressEcho} or solid-state
impurity centers~\cite{Tyryshkin03PRB,Chhildress06Science,Balasubramanian09Isotope,DuUDD} is a key issue in spin-based
quantum information processing~\cite{LaddQC}, magnetic resonance
spectroscopy~\cite{Tyryshkin03PRB,DuUDD,CoryNVdd}, and
magnetometry~\cite{MazeSensing,BalasubramanianMagnetometry,Zhao_magnetometry}. In the seminal spectral-diffusion
theories for magnetic resonance spectroscopy~\cite{AndersonSD,KuboSD}, the
couplings to the environments are taken as classical noises.
It is generally believed that different processes coupled to the same noise
source have similar decoherence behaviors and stronger noises cause
faster decoherence~\cite{AndersonSD,KuboSD}.

In modern quantum technologies, however, the relevant environments are
of nanometer size~\cite{PettaT2,KoppensT2,Greilich06Science,XuPump,Maletinsky09NP,
BrunnerHoleSpin,PressEcho,Tyryshkin03PRB,Chhildress06Science,Balasubramanian09Isotope,DuUDD,CoryNVdd} and
therefore their quantum nature becomes important. Quantum theories
developed in recent years~\cite{Witzel05PRBQuantum,Yao06Quantum,YangCCE} suggest that a
quantum nuclear spin bath, in contrast to classical noises,
possesses a great extent of controllability and has surprising
coherence recovery effects on an electron spin embedded in
it~\cite{YaoCDD}. The nuclear spin bath may also be exploited in
quantum technologies such as information storage~\cite{TaylorNMemory}.

\begin{figure}[b]
 \includegraphics[width=\linewidth]{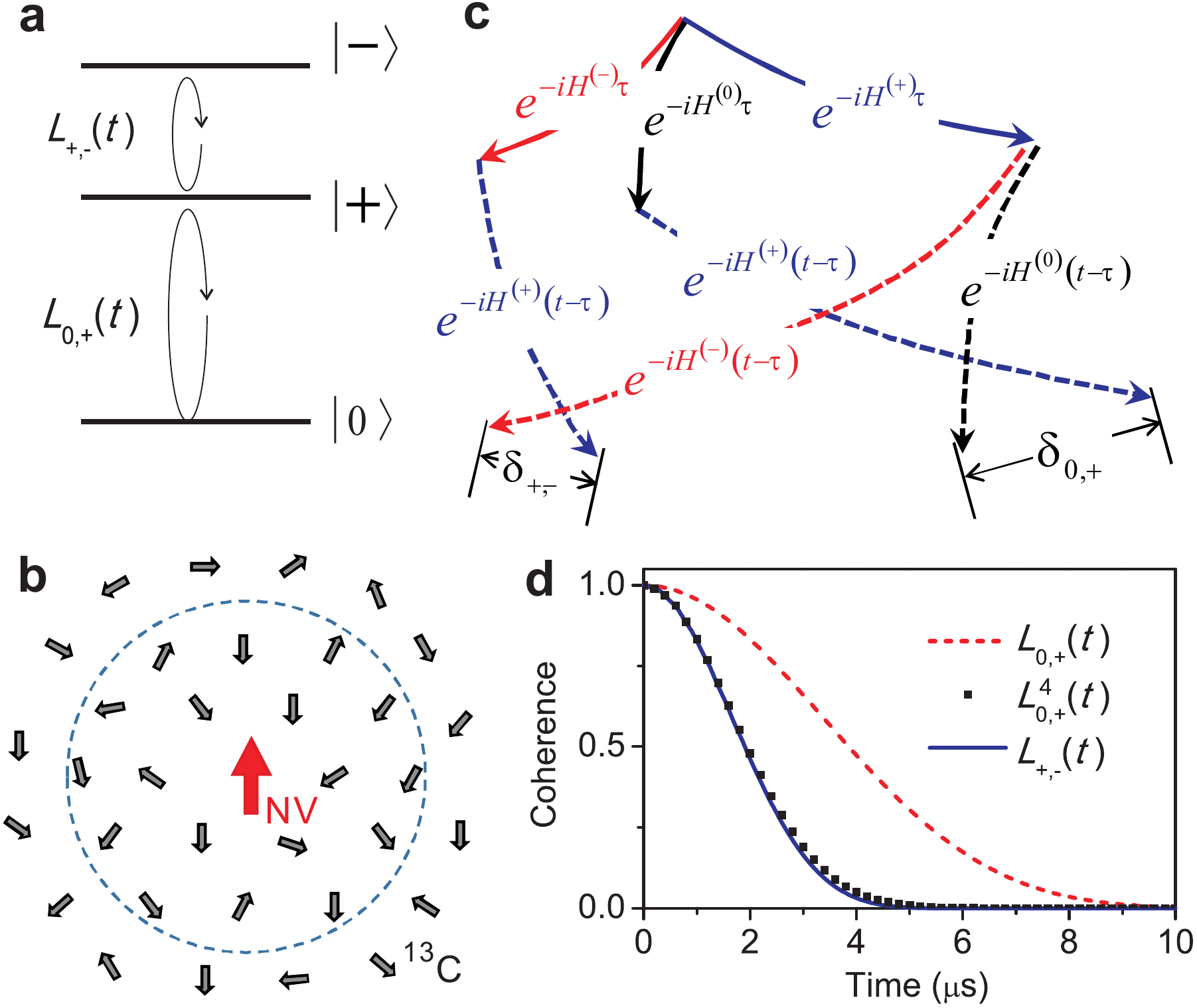}
  \caption{(Color online)  (a) The single-transition coherence $L_{0,+}(t)$ and the multi-transition coherence $L_{+,-}(t)$
  of an NV center spin.
  (b) Schematic of an NV center spin coupled to a $^{13}$C nuclear spin bath (enclosed by the circle).
  (c) Pronged quantum evolution pathways of the nuclear spin bath conditioned on the center spin states.
  The distance $\delta_{\alpha,\beta}$ (distinguishability) between the pathways
  determines the center spin coherence $L_{\alpha,\beta}(t)$. Under a flip of the center spin,
  the bath evolution directions are switched (from solid to dashed curves).
  (d) Free-induction decay of the center spin coherence $L_{0,+}(t)$ (red dashed line)
  and $L_{+,-}(t)$ (blue solid line) under a magnetic field $B=0.3$~T along the NV axis.
  The scaled single-transition coherence $L_{0,+}^{4}(t)$
  (black square symbols) is plotted for comparison.}
 \label{Fig1}
\end{figure}

In this Letter, we report an anomalous decoherence effect of a spin higher than
$1/2$ in a nuclear spin bath. We consider the multi-transition and single-transitions
of the spin-1 of a nitrogen-vacancy (NV) center in diamond [Fig.~\ref{Fig1}(a)], which are coupled to the same nuclear
spin bath [Fig.~\ref{Fig1}(b)]. Surprisingly, under the dynamical decoupling control~\cite{YangDDreview},
the multi-transition can
have longer coherence time than the single-transitions, even though
the former is subjected to stronger noises. As will be shown later, such an
anomalous effect is due to manipulation of the bath evolution via the center spin
flips. This discovery reveals a surprising aspect of quantum baths,
and paves the way for exploiting nuclear spin baths for quantum information
processing~\cite{ClarkeSCqubit,LaddQC,TaylorNMemory} and
magnetometry~\cite{MazeSensing,BalasubramanianMagnetometry,Zhao_magnetometry}.

The spin decoherence of an NV center
in high-purity (type IIa) diamond is mainly caused by
hyperfine coupling to $^{13}\text{C}$ nuclear spins~\cite{Chhildress06Science,Balasubramanian09Isotope,CoryNVdd}.
The NV center has a spin-1,
with three eigen states $|0\rangle$ and $|\pm\rangle$ quantized along the
NV ($z$) axis at zero field. The single-transitions
$|0\rangle\leftrightarrow|\pm\rangle$ and the multi-transition
$|+\rangle\leftrightarrow|-\rangle$ [Fig.~\ref{Fig1}(a)] are subjected to noises from the
same nuclear spin bath, with the noise amplitude for the latter
being twice that for the former.

In the semiclassical description, the effect on the center spin
of the environment is a fluctuating local field ${\mathbf
b}(t)$, with a Hamiltonian $H=S_{z}b_{z}(t)$, where $S_z$ has eigen states
$|0\rangle$ and $|\pm \rangle$ with eigenvalues $0$ and $\pm 1$, respectively.
Here we consider only
the fluctuations along the NV axis, since the perpendicular
components are too weak to cause spin-flip relaxation. An initial state of
the center spin
$|\Psi(0)\rangle=a_{-}|-\rangle+a_{0}|0\rangle+a_{+}|+\rangle$
will evolve to
$|\Psi(t)\rangle=a_{-}e^{i\varphi(t)}|-\rangle+a_{0}|0\rangle+a_{+}e^{-i\varphi(t)}|+\rangle$,
with an accumulated random phase
$\varphi(t)=\int_{0}^{t}b_{z}(\tau)d\tau$. The coherence of the
single-transitions $|0\rangle\leftrightarrow|\pm\rangle$ is
determined by the average of the random phase factor
$L_{0,\pm}=\langle e^{\pm i\varphi(t)}\rangle$, while the
multi-transition coherence $L_{+,-}=\langle
e^{2i\varphi(t)}\rangle$. For Gaussian noises as commonly
encountered~\cite{AndersonSD,KuboSD},
$L_{0,\pm}=e^{-\langle\varphi(t)\varphi(t)\rangle/2}$ and
$L_{+,-}=e^{-2\langle\varphi(t)\varphi(t)\rangle}$, which satisfy a
simple scaling relation
\begin{equation}
|L_{+,-}|=|L_{0,\pm}|^{4}.
\label{Eq_scaling}
\end{equation}
Decoherence of
the multi-transition behaves essentially the same as that of the
single transitions, but is faster since the multi-transition experiences
as twice strong noises as the single-transitions.

In the quantum description, the random field ${\mathbf b}$ is a quantum
operator of the bath. The bath itself has internal interaction
$H_{\rm B}$. The center spin decoherence is indeed caused by the
entanglement with the bath during the quantum evolution~\cite{Yao06Quantum}.
From the initial state
$|\Psi(0)\rangle=(a_{-}|-\rangle+a_{0}|0\rangle+a_{+}|+\rangle)\otimes|J\rangle$,
the center spin and bath evolve as
\begin{equation}
|\Psi(t)\rangle=a_{-}|-\rangle\otimes|J_{-}(t)\rangle+a_{0}|0\rangle\otimes|J_{0}(t)\rangle
+a_{+}|+\rangle\otimes|J_{+}(t)\rangle,
\label{eq:2}
\end{equation}
where $J_{\alpha}(t)\equiv\exp\left(-iH^{(\alpha)}t\right)|J\rangle$
with $H^{(\alpha)}\equiv\alpha b_{z}+H_{\rm B}$ for $\alpha=0$ or $\pm$.
The bath evolves along pronged pathways in the Hilbert space
conditioned on the center spin state [Fig.~\ref{Fig1}(c)]. The center spin
loses its coherence as its which-way information is recorded in the
bath. The coherence of the single- and multi-transitions are
determined by the overlaps between the pronged bath states as
$L_{0,\pm}(t)=\langle J_{0}(t)|J_{\pm}(t)\rangle$ and
$L_{+,-}(t)=\langle J_{+}(t)|J_{-}(t)\rangle,$ respectively. The
bath evolution can be substantially different for different center
spin states. Thus, the multi-transition may have
different decoherence behavior from what the single transitions do,
and in particular, the scaling relation in Eq.~(\ref{Eq_scaling})
does not hold in general.

A more striking difference between classical noises and quantum
baths occurs when the center spin is under the dynamical decoupling
control. In the case of classical noises, if the center spin is
under stroboscopic flips
between different states, the decoherence is controlled through
modulation of the accumulated random phase as
$\varphi(t)=\int_{0}^{t}b_{z}(\tau)F(\tau)d\tau$, where $F(\tau)$
jumps between $+1$ and $-1$ every time the center spin is
flipped~\cite{Cywinski08PRB}. In the case of quantum baths, the bath evolution
along different pathways is manipulated when the center spin is
flipped between different states. For example,
after a flip operation $|\alpha\rangle\leftrightarrow|\beta\rangle$
at time $\tau$, the electron-nuclear spin system evolves as
$a_{\beta}|\alpha\rangle\otimes e^{-iH^{(\alpha)}(t-\tau)}e^{-iH^{(\beta)}\tau}|J\rangle
+a_{\alpha}|\beta\rangle\otimes e^{-iH^{(\beta)}(t-\tau)}e^{-iH^{(\alpha)}\tau}|J\rangle$,
i.e., the bath evolutions conditioned on the center spin state exchange
their directions in the Hilbert space [Fig.~\ref{Fig1}(c)].
This results in decoherence control dramatically different from the case of
classical noises.

In our specific system, the random field results from the hyperfine
coupling to $^{13}\text{C}$ spins, $b_{z}=\sum_{j}{\mathbf
A}_{j}\cdot{\mathbf I}_{j},$ where ${\mathbf A}_{j}$ is the coupling
coefficients for the $j$th nuclear spin ${\mathbf I}_{j}$. The dipolar
hyperfine interaction decays inverse cubically with distance and the center
spin is effectively coupled to hundreds of nuclear spins located
within a few nanometers (the bath)~\cite{Chhildress06Science,Maze08PRB}.
The nuclear spins have dipolar interaction  $H_{\rm B}=\sum_{i<j}{\mathbf I}_j\cdot{\mathbb D}_{jk}\cdot{\mathbf I}_k$,
with the coupling tensor ${\mathbb D}_{jk}$ of strength about 10~Hz for two nuclei at
average distance, which is much weaker than the hyperfine coupling
($\gtrsim$~kHz for nuclei within 4~nm). During the decoherence
process, which occurs within milliseconds, negligible is the
diffusion of quantum coherence from the bath to outside. Thus, the
center spin and bath evolve as a relatively closed quantum
system [Fig.~\ref{Fig1}(b)]. An example on the contrary is an NV center in a
nitrogen-rich sample where nitrogen electron spins form the
bath~\cite{Hanson08Science,FuchsGHzNV,deLange10DD}. In that case, the interaction between two
bath spins at average distance is much stronger than that between
the center and a bath spin, and therefore the coherence diffusion in
the environment is faster than the decoherence of the center spin,
which invalidates the definition of a closed quantum bath. Instead,
the classical noise theory well describes the coupling to the
nitrogen spin bath~\cite{Hanson08Science,FuchsGHzNV,deLange10DD}.

There are also thermal noises resulting from random orientations of
the nuclear spins at finite temperature~\cite{Merkulov02PRB}, which are of
classical nature. As shown in Fig.~\ref{Fig1}(d),
the calculated free-induction decay of the single-
and multi-transition coherence, which is mainly caused by the
thermal fluctuations (also called inhomogeneous broadening) of
the $^{13}\text{C}$ nuclear spin bath, is well fitted with Gaussian decays
and satisfies the scaling relation in Eq.~(\ref{Eq_scaling}). This reflects
the classical nature of thermal fluctuations.
Indeed, the thermal fluctuations are much stronger than the quantum
fluctuations, but the inhomogeneous broadening effect can be totally
removed by spin echo. Such coexistence of classical and quantum
fluctuations, and their different effects in spin echo, can be used
for in-situ test of the semiclassical and quantum theories.

We calculate the coherence of an NV center electron spin coupled
to a nuclear spin bath that is generated by randomly
placing $^{13}\text{C}$ atoms on the diamond lattice with
natural abundance $1.1\%$. Inclusion of about 500 $^{13}\text{C}$
nuclear spins within 4~nm from the NV center is sufficient for a
converged result. For the decoherence control, we adopt
the periodic dynamical decoupling (PDD) control by
an equally spaced sequence (applied at $\tau$,
$3\tau,$ $5\tau\ldots$)~\cite{YangDDreview,CoryNVdd,deLange10DD}.

\begin{figure}[t]
 \includegraphics[width=\linewidth]{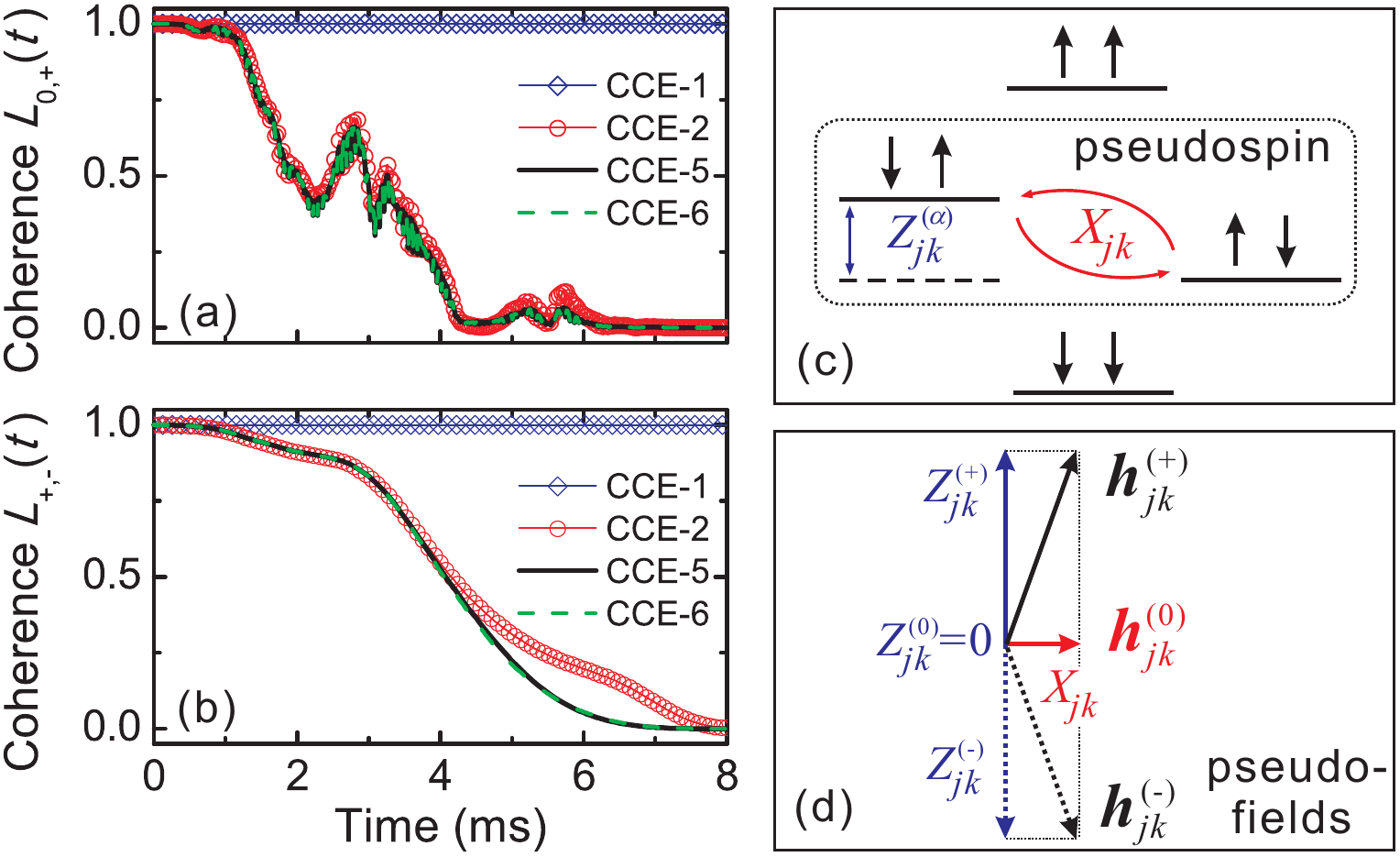}
  \caption{(Color online)
  (a) The NV center spin coherence $L_{0,+}(t)$ under PDD-5
  control and magnetic field $B=0.3$~T along the NV axis, calculated with
  different orders of CCE.
  (b) The same as (a), but for the multi-transition coherence $L_{+,-}(t)$.
  (c) The pseudo-spin picture for the nuclear spin pair dynamics.
  The polarized states $|\uparrow\uparrow \rangle$ and $|\downarrow\downarrow \rangle$
  are decoupled under a strong magnetic field. The unpolarized states
  $|\uparrow\downarrow \rangle$ and $|\downarrow\uparrow \rangle$ with
  the transition rate $X_{jk}$ and the energy difference $Z^{(\alpha)}_{jk}$
  are mapped to a pseudo-spin. (d) Schematic of the pseudo-fields
  $\mathbf{h}_{jk}^{(\alpha)}$ for different NV center spin states $|\alpha\rangle$.}
 \label{Fig2}
\end{figure}

The spin coherence is calculated with the cluster correlation expansion (CCE)
method~\cite{YangCCE}.
The center spin decoherence caused by a particular nuclear spin cluster $C$ is
denoted as $L^{(C)}_{\alpha,\beta}$. The irreducible correlation of the cluster is
recursively defined as $\tilde{L}^{(C)}_{\alpha,\beta}\equiv L^{(C)}_{\alpha,\beta}/\prod_{C' \subset C}\tilde{L}^{(C')}_{\alpha,\beta}$
which excludes the irreducible correlations of the sub-clusters $C'$.
Then the $M$th order CCE approximation (CCE-$M$) gives $L_{\alpha,\beta}\approx \prod_{|C|\le M}\tilde{L}^{(C)}_{\alpha,\beta}$ with
$|C|$ denoting the number of spins in the cluster.
In this paper, inclusion of up to 5-spin clusters (CCE-5) is sufficient
to produce converged results.

Figure~\ref{Fig2}(a) and (b) show the convergence of the CCE for the
single- and multi-transition decoherence under a strong magnetic field and
the 5-pulse PDD (PDD-5) control. The results show that
under the strong magnetic field, the single-spin dynamics in the bath (CCE-1)
is suppressed and causes negligible decoherence.
Actually, CCE-2 gives almost converged results. This means that the main mechanism
of the decoherence is the nuclear spin pair correlations.

The nuclear spin pair dynamics is essentially the flip-flop between the two states
$\left|\uparrow\downarrow\right\rangle $ and
$\left|\downarrow\uparrow\right\rangle $ [see Fig.~\ref{Fig2}(c)].
The polarized states $\left|\uparrow\uparrow\right\rangle$ and
$\left|\downarrow\downarrow\right\rangle$ are stationary,
since the nuclear spin Zeeman energy is much greater than the dipolar interaction strength.
The dipolar interaction causes the transition
$\left|\uparrow\downarrow\right\rangle \leftrightarrow
\left|\downarrow\uparrow\right\rangle $
with a rate
$X_{jk}\equiv \left\langle\downarrow\uparrow\left|
 {\mathbf I}_j\cdot{\mathbb D}_{jk}\cdot{\mathbf I}_k\right
|\uparrow\downarrow\right\rangle$.
The hyperfine interaction induces an energy cost of the flip-flop
$Z_{jk}^{\left(\alpha\right)}=\alpha\left({\mathbf A}_{j}-{\mathbf A}_{k}\right)\cdot{\mathbf e}_z$,
for the electron spin state $|\alpha\rangle$.
Thus the flip-flop is mapped~\cite{Yao06Quantum} to the precession of a pseudo-spin $\boldsymbol{\sigma}_{jk}$
about a pseudo-field $\mathbf{h}_{jk}^{(\alpha)}=\left(X_{jk},0,Z^{(\alpha)}_{jk}\right)$ [see Fig.~\ref{Fig2}(d)],
which is conditioned on the electrons spin state $|\alpha\rangle$.
The pronged bath evolution shown in Fig.~\ref{Fig1}(c), which causes the center spin decoherence,
is reduced to pronged pseudo-spin precession. The center spin decoherence caused by pair
flip-flops is factorized as
\begin{equation}
L_{\alpha,\beta}(t)\approx\prod_{jk}\left|\langle
\sigma_{jk}^{(\alpha)}(t)|\sigma_{jk}^{(\beta)}(t)\rangle\right|,
\label{eq:3}
\end{equation}
where $|\sigma_{jk}^{(\alpha/\beta)}(t)\rangle$ is the precession of the pseudo-spin
about the pseudo-field $\mathbf{h}_{jk}^{(\alpha/\beta)}$ for the center spin state $|\alpha/\beta\rangle$.

\begin{figure}[t]
 \includegraphics[width=\linewidth]{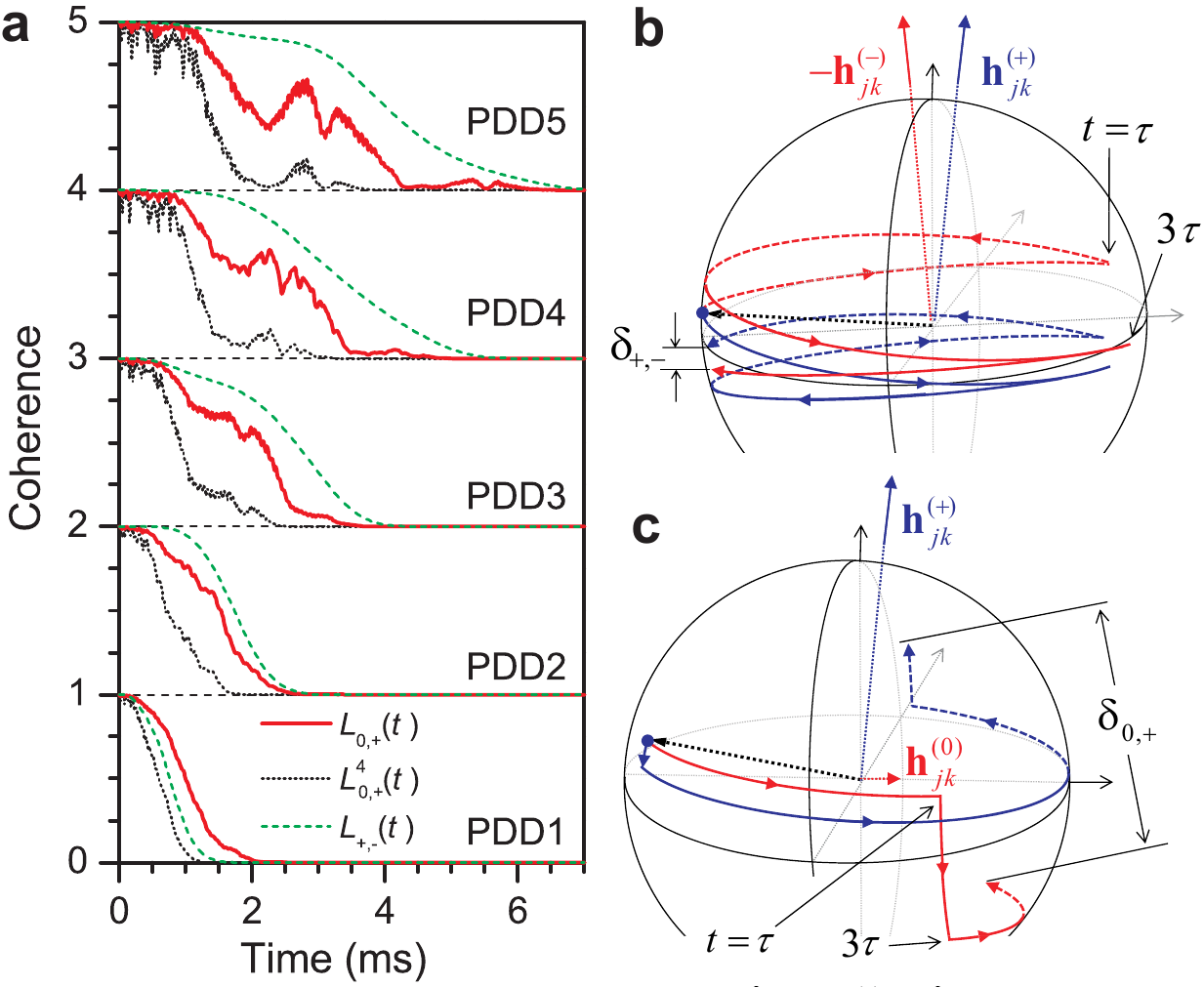}
  \caption{(Color online)
  (a) The calculated single-transition coherence $L_{0,+}(t)$ (red solid lines),
  multi-transition coherence $L_{+,-}(t)$ (green dashed lines),
  and the scaled single-transition coherence $L_{0,+}^4(t)$ (black dotted lines),
  under a magnetic field $B=0.3$~T along the NV axis and one- to five-pulse PDD control
  (PDD-1 to PDD-5, from bottom to top, vertically shifted for the sake of clarity).
  (b) The bifurcated pseudo-spin precession about the pseudo-fields
  $\mathbf{h}_{jk}^{(\pm)}$  for the multi-transition coherence under the PDD-2 control, with the initial
  state indicated by a solid circle at the end of a dotted arrow.
  Upon the center spin flip (at $t=\tau$ or $3\tau$, the pseudo-spin alternates its pseudo-field.
  (c) The same as (b), but for the single-transition coherence.
  }
 \label{Fig3}
\end{figure}

Figure~\ref{Fig3}(a) presents the main result of this paper.
Under the Hahn-echo (PDD-1) control, the inhomogeneous
broadening effect is eliminated and the decoherence is determined
by the quantum fluctuations resulting from the many-body interaction in the bath.
The multi-transition coherence decays faster than the single-transition coherence,
but the simple scaling relation in Eq.~(\ref{Eq_scaling}) is violated.
More surprisingly, when the number of control pulses is increased (from two-
to five-pulse PDD control), the multi-transition coherence even lasts longer than
the single-transition coherence.

The anomalous decoherence effect,
though counter-intuitive, can be understood using the pseudo-spin picture
as illustrated in Fig.~\ref{Fig3}(b) and (c). The center spin decoherence is determined
by the distance between the pseudo-spin pathways conditional on the electron spin states.
In the Hahn echo, the decoherence due to the pair flip-flops is~\cite{Yao06Quantum,Maze08PRB}
\begin{equation}
L_{\alpha,\beta}(2\tau)= \prod_{jk}\left[1-2\left|\sin\left({\bf h}_{jk}^{(\alpha)}\tau/2\right)
\times \sin\left({\bf h}_{jk}^{(\beta)}\tau/2\right)\right|^{2}\right].
\label{eq:4}
\end{equation}
Actually in the short time limit $h^{(\alpha)}_{jk}\tau\ll 1$, the multi-transition coherence
decays faster than the single-transition coherence and the scaling relation in Eq.~(\ref{Eq_scaling})
is satisfied. As the time increases, however, the scaling relation is violated.
For most nuclear spin pairs in the bath, the interaction between the
two $^{13}$C spins is weaker than 100~Hz, while the hyperfine energy cost of the
pairwise flip-flop is $>$kHz. Therefore in the case of multi-transition,
the two pseudo-fields ${\mathbf h}_{jk}^{(\pm)}$ corresponding to the
electron spin sates $|\pm\rangle$ are approximately anti-parallel. Thus the distance between the bifurcated
pseudo-spin pathways and hence the induced decoherence are small.
While in the single-transition case, the two pseudo-fields ${\mathbf h}^{(0)}_{jk}$ and ${\mathbf h}^{(+)}_{jk}$
are in general not (anti-)parallel, and the bifurcated pseudo-spin pathways may deviate largely from each other,
which induces strong decoherence.
As the number of PDD pulses and hence the coherence time increase, such control effect on the bath dynamics becomes significant, and therefore the
multi-transition can have longer coherence time than the single-transitions.

The conditional flip-flops of the nuclear spin pairs also explain the observation that the oscillation features in the
single-transition coherence are absent in the multi-transition coherence. A careful examination of decoherence caused
by each individual pair reveals that the rapid and shallow modulations in the decoherence profile
are induced by those pairs which have one $^{13}$C located relatively close to the NV center. Such pairs
have large hyperfine energy cost $Z^{(\pm)}_{jk}$ in the flip-flop. The large pseudo-fields cause rapid precession
of the pseudo-spin when the center spin is in the states $|\pm\rangle$ but the pseudo-spin precession for the center spin state
$|0\rangle$ is still slow. This causes rapid oscillation in the single-transition coherence.
The slow and relatively deep modulations on the single-transition coherence are caused by flip-flops of pairs which have
two $^{13}$C spins close to each other.
Neither the rapid nor slow oscillation features are visible in the multi-transition
coherence. This is because of the fact that for the multi-transition,
the two pseudo-fields ${\mathbf h}_{jk}^{(\pm)}$ are nearly anti-parallel,
and the decoherence induced by each pair is small.

In conclusion, we have discovered that the multi-transition
and single-transitions of an NV center spin in diamond, though coupled
to the same nuclear spin bath, have different
decoherence features, and more strikingly, the former can have longer coherence
time though it suffers stronger noises.
This anomalous decoherence effect establishes the quantum nature and
the controllability of an interacting nuclear spin ensemble
in a solid-state system at room temperature.

\begin{acknowledgments}
This work was supported by Hong Kong RGC/GRF CUHK402410, CUHK Focused
Investments Scheme, Hong Kong RGC HKU10/CRF/08, National Natural Science
Foundation of China Project 11028510.
\end{acknowledgments}


\end{document}